\documentclass{article}          
\usepackage{graphicx}
\usepackage{amsfonts}
\usepackage{latexsym}
\usepackage{amssymb}

\usepackage{natbib}

\def\argmax{\mathop{\rm arg\,max}\limits}
\def\nn{\nonumber}

\begin{document}

\title{Flexible parametric bootstrap for testing homogeneity against clustering and assessing the number of clusters}

\author{
Christian Hennig,\\
Department of Statistical Science,\\
University College London\\
Gower Street\\
London WC1E 6BT\\
United Kingdom\\
Tel.: +44-20-76791698\\
c.hennig@ucl.ac.uk,\\~\\
Chien-Ju Lin,\\
MRC Biostatistics Unit\\
Cambridge Institute of Public Health\\
Forvie Site\\
Robinson Way\\
Cambridge Biomedical Campus\\
Cambridge CB2 0SR\\
United Kingdom\\
chienju@mrc-bsu.cam.ac.uk
}
\maketitle

\begin{abstract}
There are two notoriously hard problems in cluster analysis, estimating the number of clusters, and checking whether the population to be clustered is not actually homogeneous. Given a dataset, a clustering method and a cluster validation index, this paper proposes to set up null models that capture structural features of the data that cannot be interpreted as indicating clustering. Artificial datasets are sampled from the null model with parameters estimated from the original dataset. This can be used for testing the null hypothesis of a homogeneous population against a clustering alternative. It can also be used to calibrate the validation index for estimating the number of clusters, by taking into account the expected distribution of the index under the null model for any given number of clusters. The approach is illustrated by three examples, involving various different clustering techniques (partitioning around medoids, hierarchical methods, a Gaussian mixture model), validation indexes (average silhouette width, prediction strength and BIC), and issues such as mixed type data, temporal and spatial autocorrelation.
~\\~\\
{\bf Keywords:} cluster validation, mixture model, distance-based clustering, Markov chain, mixed type data, spatial autocorrelation, presence-absence data\\
{\bf MSC:} 62H30, 62F03, 62F40
\end{abstract}

\section{Introduction}
Cluster analysis is about finding groups of objects in data. Cluster analysis is a key area of data analysis with applications virtually everywhere where data arise. For example, the present paper features applications in social science, biogeography and medicine. Cluster analysis methods have been developed since the 1950s in various subject areas including statistics, mathematics, computer science and machine learning, biology, psychology, and geoscience. The field of cluster analysis is therefore characterised by a lack of unification. Some cluster analysis approaches are based on probability models for each cluster, others are based on density estimation, even others are based on distance measures and discrete mathematics and do not involve probability at all. As a result, the probabilistic behaviour of cluster analysis methods is often not well understood.

In the present paper we treat two key issues in cluster analysis, namely the question whether a dataset is clustered at all, and the selection of an appropriate number of clusters. We present a general principle to address these issues, which can be applied to various approaches to cluster analysis. 

A common approach to the selection of an appropriate number of clusters $k$ is via cluster validation indexes. Cluster validation indexes are statistics that can be computed for a given clustering of a dataset and measure the quality or ``validity'' of the clustering. Various validation indexes have been proposed in the literature, for example the Calinski-Harabasz index, the Average Silhouette Width (ASW), Sugar and James's distortion, see, e.g., \cite{MiCo85,SuJa03,AGMPP12,XiongLi14}. The indexes are computed for clusterings for a range of candidate values for $k$. It is usually recommended to select the $k$ that optimises either the index or a change of the index between $k-1$ and $k$ clusters, depending on the index. These recommendations are often either purely heuristic, or based (often rather loosely) on theory using simple probability models for each cluster such as the Gaussian distribution. Some criteria for finding the number of clusters such as the Bayesian Information Criterion (BIC) for mixture model-based clustering (\cite{FrRa98}) are based on probability theory in a more consistent way, but for the purpose of the present paper they can be interpreted as cluster validity criteria as well.

We assume here that the researcher has decided which cluster analysis method and which cluster validity index to use. Our attitude regarding these decisions is that different cluster analysis methods and different validity indexes correspond to different ``cluster concepts'', which may be of interest in different applications. There is no uniquely optimal choice of a combination of these, but the researcher rather needs to decide what cluster concept is required in a specific application. For example, it may be required that all objects are, on average, represented as precisely as possible by the centroid object of the cluster to which they are assigned, which can lead, depending on the distance concept involved, to the $k$-means or the ``Partitioning Around Medoids'' (PAM) clustering method, and the Calinski and Harabasz index, see  \cite{KaRo90,CaHa74}, or the researcher may be interested in finding latent subpopulations distributed approximately according to Gaussian distributions, leading to model-based clustering with Gaussian mixtures and the BIC, \cite{FrRa98}. However, the approach taken here does not assume that clusters are in general identified with ``clustering'' probability models for data subsets such as Gaussian components of a mixture model (which do not always have characteristics that are expected of clusters such as small within-cluster distances and separation, see \cite{HeLi13}). The question whether the data can be explained by a homogeneous probability model for ``non-clustering'', or on the other hand whether there is evidence for ``real'' clustering, is treated as separate from what constitutes a cluster. The philosophy of clustering involved here has been outlined in \cite{HeLi13}.  

The main idea of the present paper is that parametric bootstrap can be used to investigate the distribution of the given validation index, simulating from a model for homogeneous data, i.e., for the absence of ``real'' clustering. The validation index can then be used as a test statistic for testing homogeneity against a clustering alternative (this yields a test for each candidate $k$ for which the index is computed, which need to be aggregated to a single homogeneity test), and the simulated null distribution can also be used to calibrate the validity index by comparing its value on the dataset against what is expected under the null model. We argue that this is a better foundation for a decision about the number of clusters than the heuristics behind the standard recommendations in most of the literature.

Although it is rarely seen in practice, the idea of setting up a hypothesis test of a null hypothesis modelling ``no clustering'' for cluster validation indexes is not new. For example it is mentioned in Chapter 4 of \cite{JaDu88}. \cite{JaDu88} mention the ``random graph'', ``random cluster label'' and ``random position'' (uniform/Poisson process distribution) hypothesis. Tests for some standard null hypotheses including the normal distribution are cited in \cite{Bock96} with a focus on a proper theoretical derivation of the distribution of the test statistics. 

Often, however, rejecting such simple null hypotheses is not evidence for clustering, because there may be more structure in the data than what these null models assume, for example temporal or spatial dependence. ``Parametric bootstrap'' refers to sampling from a parametric model with parameters estimated from the data (\cite{EfTi93}). In this paper we propose using the parametric bootstrap to sample from null models that capture the non-clustering structure in the data for testing homogeneity against clustering, and for calibrating validity indexes. The parametric bootstrap allows us to use models that are more complex and less ``theory-friendly'' than the simple models mentioned above. \cite{EfTi93} treat the parametric bootstrap somewhat briefly, because they argue that a main advantage of the bootstrap is that inference can be constructed without parametric assumptions, for which the nonparametric bootstrap, i.e., exploring the distribution of a test statistic by sampling from the observed empirical distribution, was constructed. But for the aim of testing homogeneity against clustering, the empirical distribution is not suitable, because sampling from the empirical distribution will generate datasets with the same clustering characteristic as the original dataset to be analysed. Potential homogeneity can only be explored based on a model for non-clustering. Therefore the non-parametric bootstrap is not an option here.

Using parametric bootstrap for testing homogeneity against a clustering alternative and calibration of cluster validity indexes is a very general principle. It can basically be used in every clustering problem together with any clustering method and any validation index (as long as there is enough computational power to run clustering and validation index lots of times). But every situation requires a new tailor-made null model, which means that there is no straightforward out-of-the-box way to run this approach. Readers who want to apply it need to design, implement and estimate the parameters of their own null model, capturing the structural features of their datasets that do not indicate clustering. The best way of demonstrating how to do this is to show examples. After Section \ref{sbasic}, in which the general idea is stated, it is applied to three different datasets. Section \ref{ssocial} is about mixed type data for socio-economic stratification containing continuous, ordinal and nominal variables, the latter with categories carrying somewhat stronger than purely nominal information. The clustering method is PAM and the validation index used is the ASW (\cite{KaRo90}). Section \ref{smethadone} is about a dataset giving the methadone dosages taken by patients over 180 days, involving temporal autocorrelation. PAM was applied once more, but also compared with Complete and Average Linkage clustering, and cluster validity was assessed by the Prediction Strength (PS) (\cite{TiWa05}), which explores cluster stability based on resampling. In Section \ref{ssnails} we analyse a presence-absence dataset of snail species on Aegean islands where the problem is to cluster the species distribution ranges. The null model takes into account spatial autocorrelation. Following \cite{HeHa04}, the dataset was clustered using Gaussian mixture model-based clustering with the BIC (\cite{FrRa98}) after defining a distance measure between distribution ranges and running a Multidimensional Scaling (MDS; \cite{CoCo01}). The example explores the use of the parametric bootstrap approach together with model-based clustering methods and demonstrates that the parametric bootstrap adds important information to the standard usage of model-based clustering and the BIC. Section \ref{sconc} gives a conclusing discussion.

Sections \ref{ssocial} and \ref{ssnails} give a nod to two predecessors of the current paper. \cite{HeHa04} already introduced parametric bootstrap tests for homogeneity against clustering using the specific null model that will be applied in Section \ref{ssnails}, although it did not consider the estimation of the number of clusters. The general principle proposed here was already applied in an ad hoc-fashion in \cite{HeLi13}, where the dataset of Section \ref{ssocial} was analysed. Section \ref{ssocial} improves on the null model used in \cite{HeLi13}.

\section{The general setup} \label{sbasic}
The general principle of the present paper is outlined theoretically in this section, and will then be illustrated by examples.

Given is a set of observations ${\bf X}=\{{\bf x}_1,\ldots,{\bf x}_n\}$ from some set of possible objects ${\cal X}$. The observations can be characterised in various ways, normally either by $p$ variables or by an $n \times n$-dissimilarity matrix. Then there is a clustering method $C$ so that $C({\bf X},k)=\{C_1,\ldots,C_k\}$ with $k\in K\subseteq \mathbb{N}$, and, for $i=1,\ldots,k$: $C_i\subseteq {\bf X}$. In many cases, $C$ will be a partitioning method assuming that $C_i\cap C_j=\emptyset$ for any $i\neq j$ and $\bigcup_{i=1}^k C_i={\bf X}$, and $K=\{2,\ldots,n\}$, but this is not required in general. Furthermore given is a validity index $V$, so that $V({\bf X}, C({\bf X},k))\in \mathbb{R}$ measures the quality of $C({\bf X},k)$ in some sense. We assume w.l.o.g. that a larger value of $V$ implies a better cluster quality, at least as long as clusterings with the same $k$ are compared. 

The null model ${\cal P}_0=\{P_\theta:\ \theta\in\Theta\}$ is a set of probability distributions $P_\theta$ on ${\cal X}^n$ (equipped with a suitable $\sigma$-algebra) with the interpretation that the distributions $P_\theta$ model a situation that is interpreted as homogeneous in the sense of ``absence of clustering''. The set $\Theta$ can also be rather general; in Section \ref{ssnails}, for example, $\Theta$ involves the full empirical distributions of both the sizes of species and the number of species present in the regions. Basically $\Theta$ should capture all structural information as far as it cannot be interpreted as ``clustering'', which may involve features that are usually referred to as nonparametric, such as full marginal distributions of some variables. Often the $n$ observations will be modelled as i.i.d., but this again is not required. 

Let $T_n:\ {\cal X}^n\mapsto \Theta$ be an estimator of $\theta$. For a fixed number of clusters $k\in K$ and a fixed set of observations {\bf X}, a parametric bootstrap test is defined by estimating the distribution $Q_k$, which is the distribution of $V({\bf X}, C({\bf X},k))$ under $P_{T_n({\bf X})}$, by drawing $m$ bootstrap datasets ${\bf X}_1,\ldots,{\bf X}_m$ from $P_{T_n({\bf X})}$. The bootstrapped $p$-value for testing ${\cal P}_0$ is then 
\begin{eqnarray}
 \hat p_k &=& \frac{|A_k|+1}{m+1}, \mbox{ where} \label{epk} \\
 A_k &=& \{{\bf X}_i: V({\bf X}_i, C({\bf X}_i,k))\ge  V({\bf X}, C({\bf X},k))\},
\nn
\end{eqnarray}
so that a low $\hat p_k$ implies that it is very unlikely, under $Q_k$, that $V({\bf X}^*, C({\bf X}^*,k))$ is as large or larger as the observed validity $V({\bf X}, C({\bf X},k))$, which therefore is evidence for a stronger clustering than what is expected under $Q_k$.

This defines a homogeneity test for each $k\in K$. These need to be aggregated into a single homogeneity test. This can be done by defining an overall aggregated $p$-value
\begin{eqnarray}
 \hat p &=& \frac{|A^*|+1}{m+1},  \mbox{ where} \label{ep} \\
A^* &=& \left\{{\ bf X}_i:\ \sum_{k\in K}\tilde p_k({\bf X}_i)\le \sum_{k\in K}\hat p_k\right\}, \nn 
\end{eqnarray}
where $\tilde p_k({\bf X}_i)$ is the analogue of $\hat p_k$ based on\\ $V({\bf X}_i, C({\bf X}_i,k))$, i.e., with ${\bf X}_{m+1}={\bf X}$:
\begin{eqnarray*}
& \tilde p_k({\bf X}_i) = \frac{|\tilde A_i|+1}{m+1}, \mbox{ where} &\\
& \tilde A_i =
\{{\bf X}_j:\ j\neq i,\ V({\bf X}_j, C({\bf X}_j,k))\ge  V({\bf X}_i, C({\bf X}_i,k)) \}.&
\end{eqnarray*}
This means that aggregating the tests for all $k\in K$ is effectively based on averaging the $p$-values or, equivalently, ranks of $V({\bf X}, C({\bf X},k))$ among the bootstrapped samples over $k$.

An optimal value of $k$ can be found by maximising a calibrated $V$:
\begin{eqnarray}\label{ek}
\hat k&=&\argmax_{k\in K}\frac{V({\bf X}, C({\bf X},k))-EV_k}{SV_k},\\
EV_k &=& \frac{1}{m}\sum_{i=1}^m V({\bf X}_i, C({\bf X}_i,k)),\nn\\
SV_k &=& \sqrt{\frac{1}{m-1}\sum_{i=1}^m\left(V({\bf X}_i, C({\bf X}_i,k))-EV_k\right)^2}.
\nn
\end{eqnarray}
The interpretation is that this is the $k$ for which\\ 
$V({\bf X}, C({\bf X},k))$ gives the best validity compared to what 
is expected under $Q_k$.

All the information from the parametric bootstrap can be visualised by plotting 
$V({\bf X}, C({\bf X},k))$ together with all $V({\bf X}_j, C({\bf X}_j,k))$ against $k$ (``bootstrap validity plot''), which will be done in the following sections. Actually, often this plot will be so expressive that computing the formal outcomes (\ref{epk}), (\ref{ep}) and (\ref{ek}) does not add much information. 

There are alternative ways to define the tests and the estimation of $k$ based on the parametric bootstrap. Instead of (\ref{epk}), $V({\bf X}, C({\bf X},k))$ could be standardised as in (\ref{ek}), and the $p$-value could then be computed from a Gaussian distribution, although the Gaussian approximation cannot be proved to work in the generality required here. Instead of averaging $p$-values in (\ref{ep}), one could also average raw values of $V$ (implicitly assuming that these are meaningfully comparable over $k\in K$), or one could use a Bonferroni-adjustment of the lowest $\hat p_k,\ k\in K$, which can be very conservative but may work well if $V({\bf X}, C({\bf X},k))$ for the best $k$ is expected to stand out clearly. A comparison of these options is left to future work.

The parametric bootstrap tests proposed here require the specification of a null model, but they do not require the explicit specification of alternative models. The ``clustering alternative'', against which the homogeneity null model is tested, is implicitly defined by the choice of $V$. The ``effective alternative'' are distributions $P$ on ${\cal X}^n$ under which the distribution of $V$ is stochastically larger than under ${\cal P}_0$. The choice of $V$ therefore defines the meaning of ``clustering'' against which the homogeneity null hypothesis is tested.

A general limitation of parametric bootstrap is that the distribution $P_{T_n({\bf X})}$ is usually interpreted to represent the whole of ${\cal P}_0$. $\hat p$ will be anti-conservative as a $p$-value for testing ${\cal P}_0$ to the extent that other distributions in ${\cal P}_0$ exist that are both compatible with the observed data and tend to deliver larger values of $V$. Theoretical analysis of this problem is impossible in general and probably very tedious if possible at all in most specific situations and will therefore not be done here. The validity of significant test results therefore relies on the quality of the estimator $T_n$ and the assumption that different values of $\theta$ (at least as long as they are still compatible with the data) do not tend to yield vastly different values of $V$.

\section{Socio-economic stratification (mixed type data)} \label{ssocial}
\subsection{Data}
\cite{HeLi13} analysed a dataset from the 2007 US Survey of Consumer Finances. There were $n=17,430$ individuals and 8 variables (no missing values were in the dataset):
\begin{itemize}
\item $\log(x+50)$ of total amount of savings as of the average of
last month (treated as continuous),
\item $\log(x+50)$ of total income of 2006 (treated as continuous),
\item years of education between 0 and 17; this is treated as ordinal
(level 17 means ``graduate school and above''),
\item number of checking accounts that one has; this is ordinal with
6 levels (corresponding to no/1/2/3/(4 or 5)/(6 or more) accounts,
\item number of savings accounts, coded as above,
\item whether or not one has life insurance (binary, i.e., ordinal),
\item housing, nominal with 9 levels:
``neither owns nor
rents'', ``inapplicable'',
``owns or is buying/land
contract'', ``pays rent'', ``condo'', ``co-op'', ``townhouse association'',
``retirement lifetime tenancy'' and ``own only part'',
\item occupation class, nominal with 7 levels (from 0 to 6):  
``not working for pay'', ``managerials and professionals'', 
``white collar workers and technicians'', 
``lower-level managerials and various professions'',
``service workers and operators'', ``manual workers and operators'',
``farm and animal workers''. 
\end{itemize}
The aim was to use clustering methods for socio-economic stratification (see \cite{HeLi13} for background). An interesting issue, which is addressed by the approach of the present paper, is whether such a stratification is rather an artificial (although potentially useful) partition of a rather homogeneous population, in which there are no clear boundaries between social classes or strata (note that the term ``homogeneous'' here does not refer to social equality).  

\subsection{Clustering method and validation index}
\cite{HeLi13} settled for the PAM clustering method (namely for its large sample version CLARA) and the ASW validation index (\cite{KaRo90}), arguing that social strata should be defined by low within-cluster distances rather than components of a mixture model. 

PAM tries to find $k$ centroid objects in ${\bf X}$ and assigns all objects in ${\bf X}$ to the closest centroid so that the sum of distances of every object to its cluster's centroid is minimised. The ASW averages standardised differences between the average distance of every object to the closest cluster to which it is not classified and the average distance to all objects of the cluster to which it is classified. The ASW can be between -1 and 1. Values larger than 0 indicate that objects have on average lower distances within their own cluster than to their neighbouring cluster, which can be seen as minimum requirement for a distance-based clustering. Larger values of the ASW are better and \cite{KaRo90} recommend to maximise the ASW for finding the best value of $k$, but this is not based on an analysis of what changes of the ASW can be expected when increasing $k$. This can be explored by the parametric bootstrap approach. 

The distance measure was in principle a Euclidean distance for which the ordinal variables were used with standard Likert coding (1, 2, 3, \ldots) and the nominal variables were coded by binary dummies for the categories. However, variables were standardised with specific standardisation schemes in order to balance the contribution of the different types of variables to the clustering in an appropriate way. The account number variables were weighted down because they were comparably less important than the others, and the dummy variables for housing were weighted is such a way that the effective distance treated ``owns'' and ``pays rent'' as the extremes of the scale with the other categories in between, see \cite{HeLi13} for details. 

\subsection{Non-clustering structure}
For running the parametric bootstrap, a null model needs to be defined that captures the non-clustering structure in the data. This requires some judgement by the researchers, because it depends on what constitutes a ``significant clustering'' in the given application. Before defining the null model, here is some discussion of what the non-clustering structure is.

A first thing to realise is that dependence between variables may lead certain clustering methods into building clusters with approximate local independence (i.e., independent within clusters). For standard latent class clustering of nominal variables, local independence is a standard assumption (e.g.,\cite{HeLi13}), as well as for $k$-means clustering if this is written down as a Maximum Likelihood (ML)-method for spherical Gaussian clusters. PAM and the ASW do not formally assume local independence. But if a number of variables are strongly correlated with a continuous transition without clear cluster boundaries between low values on all variables and high values on all variables on at least a subset of the observations, this subset does not make a suitable cluster according to most distance-based clustering methods including PAM/ASW, because there would be very large distances within the cluster between objects that are low or high, respectively, on all variables. Researchers may well be interested in splitting up such subsets for practical purposes such as information reduction, but the resulting clustering may not be interpreted as ``real'' in the sense that there is no separation and therefore no ``natural'' cluster boundaries. Therefore, dependence between variables is seen as non-clustering structure here.

Furthermore, the information about the categories of the variables housing and occupation is somewhat stronger than nominal, which may cause some structure in the data which does not contribute to clustering that is interpretable as ``real''.

The modelling of the marginal distributions of the variables is a subtle issue. We regard the marginal distribution of nominal variables as not carrying clustering information, because a low relative frequency of certain categories cannot be interpreted as a cluster-defining ``gap'' between other categories. Therefore, the null model should reproduce the marginal distribution of the nominal variables as ``non-clustering structure''. But the situation is different for continuous distributions. The marginal distribution of a continuous variable can have various modes and gaps between them, which can be taken as indicating ``real'' clustering boundaries indeed. This means that a model is needed for a marginal distribution of the continuous variables which can be interpreted as non-clustering. Ordinality means that there is no metric distance between the categories, which indicates that low frequency-categories should not be interpreted as ``gap between clusters'', similar to the situation for nominal variables. Therefore we will treat the marginal distribution of ordinal variables as non-clustering structure and therefore as a parameter. 

In \cite{HeLi13}, the Gaussian distribution was used as marginal null distribution for the continuous variables. A Gaussian distribution can properly be interpreted as ``non-clustering'', but it may be too restricted. The null model may be rejected not because there is a real clustering, but because the real marginal distribution is non-normal, for example skew. A more flexible way of modelling non-clustering structure is to use a general unimodal distribution. Furthermore, a special feature of the continuous variables is that there are a number of individuals with zero savings (7434) and/or zero income (5). Particularly the large group of zero savings individuals causes strong non-normality of the marginal distribution. It is a matter of judgement whether such a group of individuals sharing the same value on a variable alone is seen as indicating a real clustering or not. For the present paper, we decided that we do not want to interpret this as indicating clustering, and therefore it needs to be incorporated in the null model. Note that using a Gaussian distribution as the null marginal as in \cite{HeLi13} implicitly amounts to interpreting this deviation from a Gaussian as evidence for real clustering. Obviously the existence of a large number of Americans with zero savings is something real; whether this is interpreted as ``clustering'', though, depends on whether one imagines the existence of such a group in a ``classless'' society. Here we take the point of view that even in such an idealised homogeneous classless society, people are still free to spend all their savings and a considerable number of them will do that.

\subsection{Null model} \label{snm1}
\begin{figure*}
\begin{center}
  \includegraphics[width=0.7\textwidth]{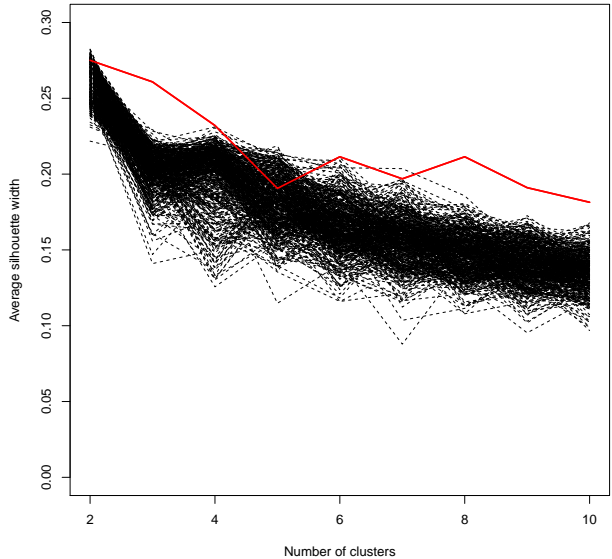}
\end{center}
\caption{ASW values for socio-economic stratification data (red) and
500 bootstrap samples (black) for $k=2,\ldots,10$.}
\label{fsocial}       
\end{figure*}

The null model has to be defined in such a way that its parameters can be estimated by the data. Some aspects of the estimation will be ad hoc so that it is not needed to set up a full new estimation theory. It is therefore useful to think about the definition of the null model and the estimation of the parameters in one go.

We choose 
${\cal P}_0$ to be based on a latent Gaussian model, the outcomes of which are transformed to match the marginal distributions mentioned above. The correlations of such a model can be estimated from ordinal data by using the technique of polychoric correlations (\cite{Dra86}). 

Assume ${\bf x}_1,\ldots,{\bf x}_n$ to be i.i.d. 
distributed, with ${\bf x}_1=(x_{11},\dots,x_{1p})^t$ generated from a 
latent $p$-variate ($p=8$) Gaussian random variable 
${\bf z}=(z_1,\ldots,z_p)^t\sim {\cal N}({\bf 0}_p, {\bf \Sigma})$, 
${\bf \Sigma}$ being a correlation matrix, i.e., with diagonals equal to one. 
For the continuous variables ($j=1, 2$) assume 
$P\{x_{1j}=\log(50)\}=p_j>0$ (remember that the continuous variables are 
$\log(50+x)$-transformed, so $\log(50)$ refers to zero savings or income) 
and the conditional distribution 
${\cal L}(x_{1j}|x_{1j}>\log(50))$ to be unimodal with continuous density. Let 
$G_j$ be the cdf of the full distribution of $x_{1j}$, and assume that
$x_{1j}=G_j^{-1}(\Phi(z_j))$, where $\Phi$ is the cdf of the standard Gaussian
distribution.

For an ordinal variable ($j>2$; see below for nominal variables)  
$x_{1j}$ with $h$ categories $c_1,\ldots, c_h$ let 
$-\infty=u_{j0}<u_{j1}<\ldots<u_{jh}=\infty$ be a sequence of Gaussian quantiles so 
that $x_{ij}=c_g \Leftrightarrow z_j \in (u_{j(g-1)},u_{jg}]$, $g=1,\ldots,h$.

Furthermore, assume that the categories of the nominal variables ($j=8, 9$)
$x_{1j}$
are ordered with an unknown ordering, and the true ordering is defined 
according to the average 
true correlations between the dummy variables indicating the categories of
$x_{1j}$ and the variables that were originally ordinal or continuous.
Given this true ordering,
$z_j$ is related to $x_{ij}$ as for the ordinal variables above.

\subsection{Null model parameter estimation} 
In order to estimate the polychoric correlations, i.e., the matrix ${\bf \Sigma}$, the continuous variables are treated as ordinal by splitting them up into 10 ordered categories each, of approximately the same size. The true orders of the nominal variables ($j=8, 9$) can be estimated by computing the average sample correlations between the dummy variables indicating the categories of
$x_{1j}$ and the variables that were originally ordinal or continuous. With that, all the variables are ordinal and ${\bf \Sigma}$ can be estimated as in 
\cite{Dra86}. $-\infty=u_{j0}<u_{j1}<\ldots<u_{jh}=\infty$ can be estimated so that they reproduce the observed marginal distributions of the ordinal and nominal variables.

The distributions $G_j$ can be estimated by using the empirical probability for $x_{1j}=\log(50)$, and by fitting a kernel density estimator to the observations with $x_{ij}>\log(50)$ making sure that the estimated density is unimodal. In practice this has been done by using the ``density'' function in R with default settings, and if this was not unimodal, by increasing the bandwith by steps of the originally selected bandwidth divided by 20 until the resulting density is unimodal.

\subsection{Parametric bootstrap}
In order to explore the distributions ~\\
$Q_k=P_{T_n({\bf X})}(V(\bullet, C(\bullet,k)))$, repeat $m$ times:
\begin{enumerate}
\item Generate $n$ i.i.d. observations ${\bf z}_1^*,\ldots,{\bf z}_n^*$ 
from\\ ${\cal N}({\bf 0}_p, \hat{\bf \Sigma})$.
\item Transform them into ${\bf X}^*=\{{\bf x}_1^*,\ldots,{\bf x}_n^*\}$ according to
Section \ref{snm1}, using the estimated distributions $\hat G_j$ and Gaussian
quantiles $\hat u_{jg}$.  
\item Compute a distance matrix for the objects in ${\bf X}^*$ as in \cite{HeLi13}.
\item For $k\in K$, cluster ${\bf X}^*$ by PAM and compute and store the ASW, 
$V_{qk}=V({\bf X}^*, C({\bf X}^*,k))$,\\ $q=1,\ldots,m$.
\end{enumerate}

\subsection{Results}

The results with $m=500$ and $K=\{2,\ldots,10\}$ are shown in Figure 
\ref{fsocial}. The real dataset produces clearly
outstanding ASW values overall except for $k=2$ and $k=5$. The average 
$\hat p_k$-value is smaller than for all datasets generated from the
null model, so  $\hat p=\frac{1}{501}$, the smallest possible value, a
strong rejection of the homogeneity model.

For $k=2$ the raw ASW reaches its maximum, so according to the standard 
recommendation
$k=2$ should yield the best clustering. But for $k=2$ the real dataset
does not yield a significantly better clustering than the null model, as
opposed to most other values of $k$. This means that for this dataset
the standard recommendation is misleading. A higher $k$ gives a better
ASW in comparison to what can be expected under the null model. The best 
calibrated ASW values as in (\ref{ek}) are 4.476 for $k=3$ and 4.481
for $k=8$, which leads to $\hat k=8$ as recommended value. $k=3$ is about 
as good.

In \cite{HeLi13}, where the Gaussian distribution was used as the null marginal for the continuous variables, the conclusions were mainly the same, but the null model here improved the achieved ASW-values to some extent, with the effect that now the real dataset no longer produces the largest ASW-value for every single $k>2$. This shows that assessment in \cite{HeLi13} was somewhat over-optimistic regarding the strength of the clustering, but the main conclusion is confirmed.

\section{Methadone patients (Markov time series)} \label{smethadone}
\subsection{Data}

\begin{figure}
\begin{center}
  \includegraphics[width=0.5\textwidth]{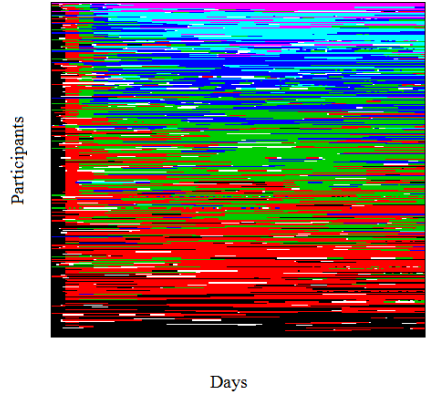}
\end{center}
\caption{Heatplot of methadone data. Dosage categories are 1 (black), 2 (red), 
3 (green), 4 (blue), 5 (light blue), 6 (violet), missing (white). Patients are ordered according to average dosage.}
\label{fmethadone}       
\end{figure}

\cite{Lin14} analysed dosage pattern data from 314 Taiwanese 
heroine addicts receiving methadone. For every methadone patient, the dataset
contains records of the methadone dosage taken on each of the first 180 days from the beginning of the methadone therapy. There are six ordered dosage categories 1-6. Also there are missing values, meaning that the patient did not show up for obtaining methadone on a certain day. The data are shown in Figure \ref{fmethadone}. Cluster analysis was done partly exploratory and partly for making the communication about the dosage patterns simpler. A clustering therefore can be useful regardless of whether there is some real clustering pattern in the data or not, but a really meaningful clustering, if it exists, is of medical interest. 

Again, clusters should be characterised by small distances within clusters.
\cite{Lin14,LiHeHu15} defined a distance measure in which missing values were treated as an additional category not having ordinal information, and the other categories were treated, following expert advice, in such a way that a change from one category to any other category was treated as fairly substantial, even between neighbouring categories, which means that categories were treated as carrying some compromise between ordinal and nominal information. 

\subsection{Clustering method and validation index}\label{smethodmethadone}
Several clustering methods were compared, namely PAM, Average Linkage and Complete Linkage hierarchical clustering (\cite{KaRo90}). Cluster validation was done by two methods, namely the ASW and the PS (\cite{TiWa05}). We focus on the latter one here. The PS measures the stability of the clustering. The idea of the PS is as follows: The dataset is split into two equally sized parts $b$ times. Every time, both halves are clustered into $k$ clusters by the method that is also applied to the original dataset. For each of the two clusterings computed on the two halves, a prediction rule is created for the observations of the other half of the data. For any pair of observations in the same cluster in the same half it is then checked whether or not they are predicted into the same clustering by the clustering on the other half. If this is the case, their co-membership was correctly predicted. The PS is defined by averaging the proportions of correctly predicted co-memberships for the weakest clusters in each of the $2b$ halves. The prediction rule recommended in \cite{TiWa05} is to predict observations into the cluster with the closest cluster mean, which is appropriate for $k$-means. For PAM, we chose the closest centroid, for Average Linkage the minimum average distance and for Complete Linkage the minimax distance. 

The PS is not calibrated for properly comparing values for different $k$; it can be expected that larger values of $k$ make it more difficult to achieve a high PS, particularly because the PS is determined by the least stable cluster. Because of this, \cite{TiWa05} suggest to choose the largest $k$ for which the PS is larger than 0.8. Instead, because PS-values for fixed $k$ are comparable, the parametric bootstrap idea can be applied. This amounts to $b$ sample splits for each $k$ and each of the $m$ bootstrap samples, which makes this very computer-intensive.

\subsection{Non-clustering structure}
Here is some background knowledge about the methadone dosages. Obviously, for a given patient, the different days cannot be treated as independent. Once a week, every seven days, the methadone patients get a new prescription. On all other days, the patients are free to use a smaller dosage than indicated on their prescription, so that there are occasional changes on these days, but most changes happen on day 1, 8, 15 etc. (``prescription days''). Some patients go to two different doctors and obtain two different prescriptions, which makes their dosages more flexible. The dataset does not include prescription data; we only know what dosage the patient took, but not what the prescription was; so it is not possible to use the prescription for setting up the null model. In any case, even on prescription days, old prescriptions are often renewed, and if there is change, it is mostly by only one category. There is much more change on the earliest prescription days when doctors do not yet have much experience with the patients than later. Furthermore, most patients start on the lowest dosage. There is no obvious connection between previous dosages and missing values; previous missing value behaviour is much more informative about missing values in the future than are observed dosages.

\subsection{Null model} 
Modelling the time series of dosages of a single patient as a Markov chain ignores long range dependence. We choose this approach anyway, because other features of the data are more striking, and a Markov chain with different transition probabilities for prescription days and ``normal days''  already requires a large number of parameters. Some transition probabilities, particularly between non-neighbouring dosages, are very small and difficult to estimate with the limited amount of data.

The dosages of the $n$ patients (ignoring missing values for the moment) are modelled in ${\cal P}_0$ as i.i.d. Markov chains with different transition probability matrices between the six dosages for a) prescription day 2 (prescription day 1 defines the initial distribution of dosages), b) prescription day 3, c) all further prescription days combined, and d) all normal days combined. Furthermore we assume an unrestricted distribution of missingness patterns (i.e., a series of indicators of whether a patient is missing or not over all 180 days), from which one is drawn i.i.d. for each patient, independently of the patient's dosages. This effectively rules out missingness patterns as sources for ``real'' clustering; whatever is observed can be reproduced by the null model. This seems appropriate because real clusterings are only of medical use if they correspond to patterns of dosages. Missingness does not allow an interpretation in terms of the methadone needs of a patient.

\subsection{Null model parameter estimation} 
The four transition probability matrices can be estimated in a straightforward manner by empirical transition probabilities in the given situations. The initial dosage of a patient can be drawn randomly from the empirical distribution of initial dosages. The distribution of missingness patterns can also be estimated directly as its empirical distribution.

\subsection{Parametric bootstrap}
Repeat $m$ times:
\begin{enumerate}
\item For $n$ i.i.d. observations in the bootstrap sample ${\bf X}^*$:
  \begin{enumerate}
  \item Draw an initial dosage from their empirical distribution.
  \item Generate a sequence of 180 dosages using on each day the appropriate estimated transition probabilities.
  \item Independently of the sequence of dosages, draw a missingness pattern (i.e. a set of days with missing values) from the empirical distribution of missingness patterns, and make the corresponding days missing.
  \end{enumerate}
\item Compute a distance matrix for ${\bf X}^*$ as explained in \cite{Lin14}.
\item Repeat $b$ times:
  \begin{enumerate}
  \item Split ${\bf X}^*$ into two equally large subsets.
  \item Cluster both subsets by PAM for $k\in K$, and by Average Linkage and
  Complete Linkage; for the latter two methods, clusterings for $k\in K$ can be obtained by cutting the 
  trees at the appropriate height.
  \item For all pairs of observations within clusters, check whether the co-memberships are correctly predicted as explained in Section \ref{smethodmethadone}. 
  \end{enumerate}
  \item For the $2b$ clusterings, the three clustering methods, and $k\in K$, compute and store the PS, i.e., the minimum (over the $k$ clusters) proportion of correctly predicted co-memberships in the cluster. Average these to get 
$V_{qk}=V({\bf X}^*, C({\bf X}^*,k))$, $q=1,\ldots,m$.
\end{enumerate}

\subsection{Results}

\begin{figure}
\begin{center}
  \includegraphics[width=0.45\textwidth]{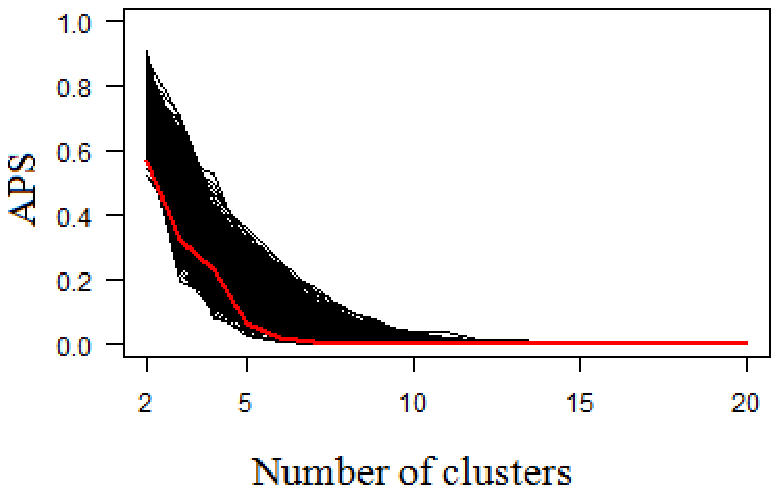}
  \includegraphics[width=0.45\textwidth]{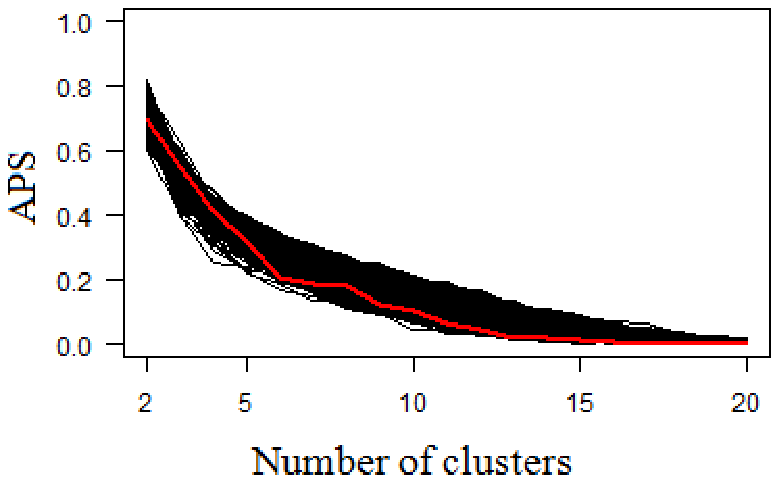}
\end{center}
\caption{PS values for methadone data (red) and
500 bootstrap samples (black) for $k=2,\ldots,20$, Average Linkage clustering
(above), Complete Linkage clustering (below)}
\label{fmethalcl}       
\end{figure}

\begin{figure}
\begin{center}
  \includegraphics[width=0.45\textwidth]{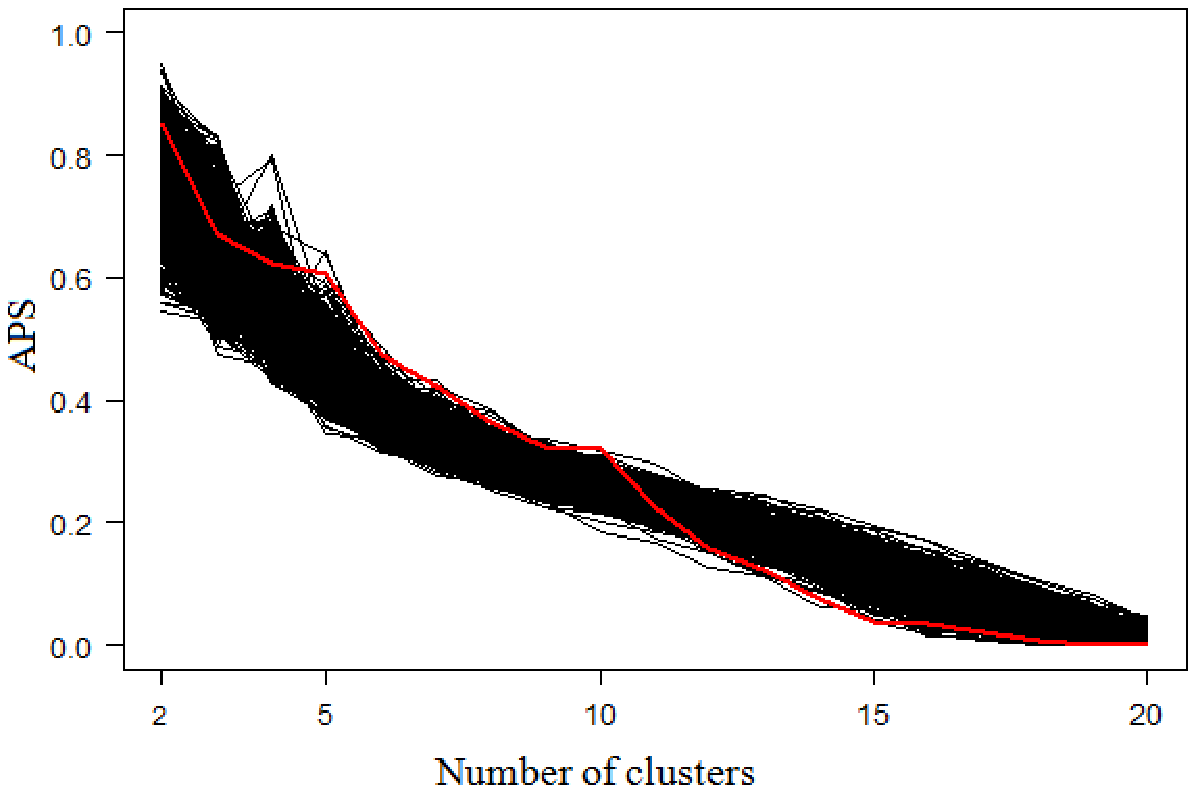}
  \includegraphics[width=0.45\textwidth]{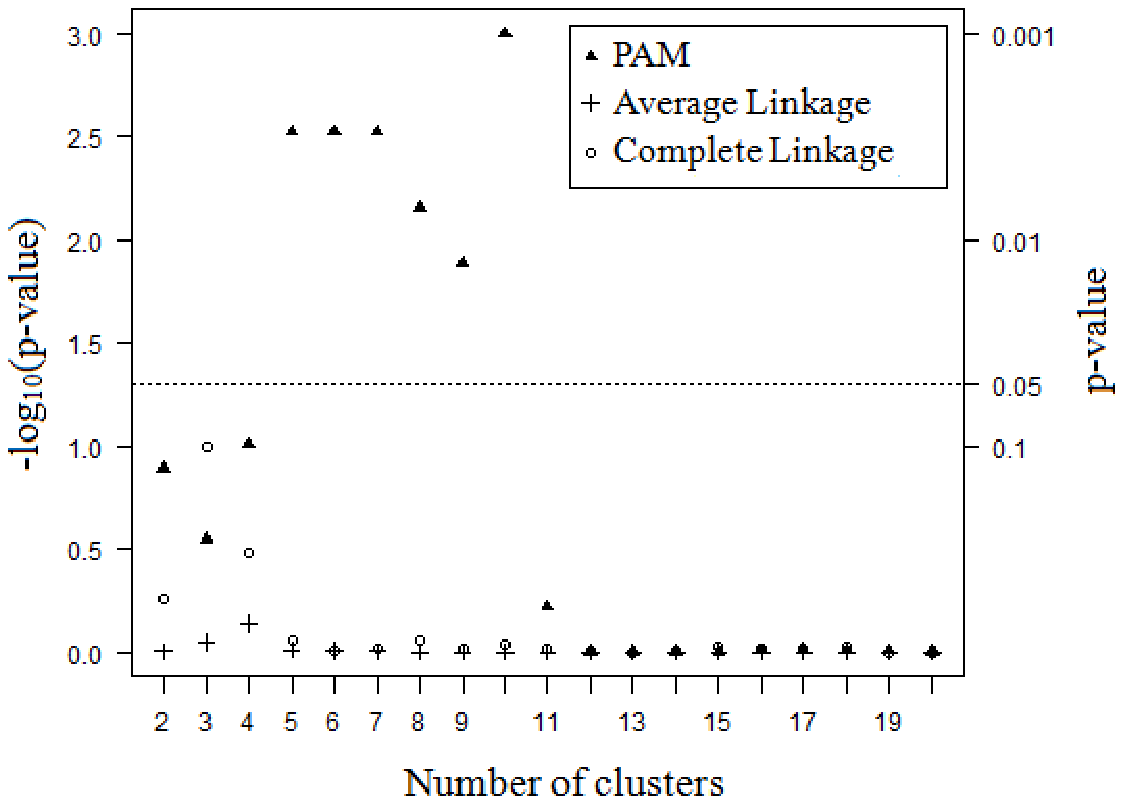}
\end{center}
\caption{PS values for methadone data (red) and
500 bootstrap samples (black) for $k=2,\ldots,20$, PAM clustering
(above); bootstrapped p-values 
$\hat p_k$ for all three clustering methods (below)}
\label{fmethpam}       
\end{figure}

The results based on $m=500,\ b=50$ are shown in Figures \ref{fmethalcl} and \ref{fmethpam}. For Average Linkage and Complete Linkage, the original methadone dataset yields stability (PS) values that are on average even below the values from the null model. Obviously, based on these clustering methods, there is no evidence for real clustering. For PAM, looking at some values of $k$, namely $k\in\{5,\ldots,10\}$ the PS values look significantly higher than those from the null model (with $k=10$ looking best; see the right side of Figure \ref{fmethpam}), although they do not clearly stick out. However, avoiding cherry-picking the best values of $k$, averaging $\hat p_k$ values over $k$ according to (\ref{ep}) gives $\hat p=0.475$ due to some very low PS-values for higher $k$, and therefore here again there is no evidence for real clustering (strictly speaking, having tried out three clustering methods, even a further adjustment for multiple testing would be required). 

Once more, the default recommendation, which for the PS is ``take the largest $k$ for which PS$>0.8$'', turns out to be misleading. The real dataset achieves this for PAM and $k=2$ only, but this does not seem to be the best value compared to the null model, and PS$>0.8$ can be achieved by the null model for $k=2$ for all clustering methods, and for PAM occasionally even for $k=3$.

\section{Distribution ranges of snail species (spatial dependence)}  \label{ssnails}
\subsection{Data}
The third example dataset is a binary dataset giving presence-absence information on 80 species of snails for 34 Aegean islands (Cyclades; \cite{HaHe05}). The dataset is available under the name ``kykladspecreg'' in the R-package ``prabclus''. Clustering of such species distribution ranges aims at finding ``biotic elements'', groups of species sharing specific areas, which are connected to certain hypotheses about speciation (\cite{HaHe03}). We here interpret the observations ${\bf x}_i,\ i=1,\ldots,n$  as sets of islands for which the species is present.

\subsection{Clustering method and validation index}
Following \cite{HaHe03} and \cite{HeHa04}, Kulczynski-dissimilarities were computed between different species:
\begin{displaymath}
d({\bf x}_1,{\bf x}_2)=1-\frac{1}{2}\left(\frac{|{\bf x}_1 \cap {\bf x}_2|}{|{\bf x}_1|}+ \frac{|{\bf x}_1 \cap {\bf x}_2|}{|{\bf x}_2|}\right).
\end{displaymath}
Classical MDS (\cite{CoCo01}) was used to map the species onto 4-dimensional Euclidean space, and the resulting points were clustered by fitting a Gaussian mixture model with a uniform noise component using the R-package ``mclust'' (\cite{FrRa98,FrRa02,FRMS12}). The reason for this was that experience with such datasets suggested that biotic elements may differ quite a bit regarding within-cluster variation, which could be captured better with the potentially different covariance matrices of Gaussian distributions than with standard distance-based clustering methods, see \cite{HeHa04} for details.

A standard way to estimate the number of clusters (and also potential constraints of the covariance matrices) is the BIC, which here is defined as $2l_n(k)-r(k)\log(n)$ so that large values of the BIC are good, where $l_n(k)$ is the log maximized likelihood for $k$ Gaussian components and $r(k)$ is the corresponding number of free parameters. The BIC is the default method in ``mclust''. It can be computed for $k=1$ and it can therefore also deliver a decision about whether $k=1$ (interpreted as homogeneity of the distribution or absence of biotic elements) or not. The BIC is here used as cluster validation index, although this is not how it would normally be interpreted, because it is defined based on probability theory within a certain model. In Section \ref{spars} a modification is used, which compares the BIC value for $k>1$ with the one for $k=1$.

\subsection{Results with plain Gaussian null model}\label{splains}

\begin{figure}
\begin{center}
  \includegraphics[width=0.5\textwidth]{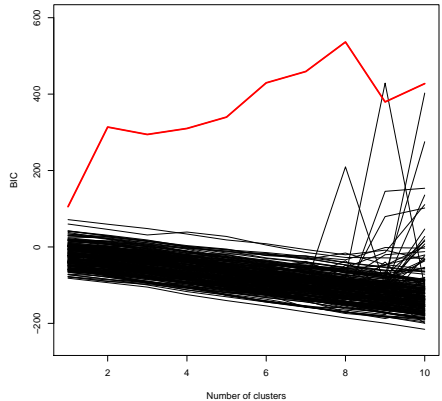}
\end{center}
\caption{BIC values for Aegean islands snails data (red) and
500 bootstrap samples (black) drawn from plain Gaussian null model 
for $k=1,\ldots,10$.}
\label{fbicnormal}       
\end{figure}

Given that the BIC is proved to be consistent under certain (somewhat restrictive) conditions (\cite{Ker00}), and that it makes a decision involving a homogeneous ``null model'', one may wonder whether parametric bootstrap adds anything to using the Gaussian mixture model combined with the BIC. Figure \ref{fbicnormal} was produced by parametric bootstrap with bootstrap data generated from a simple Gaussian null model, parameters of which were estimated in the standard way from the MDS output ($m=200$). These data were clustered by ``mclust'' in the same way as the original snails data, and the BIC was computed for $k=1,\ldots,10$.  
The BIC points to 8 clusters. Given that the expected BIC seems to go down linearly under the null model (it should indeed yielld a maximum for $k=1$, which is true under the null model), the clustering indeed seems to be highly significant, and the maximum of the BIC at $k=8$ seems to be confirmed as optimal number of clusters. The only new thing that the parametric bootstrap shows is that the null model produces occasional outliers (caused by spurious clusters that generate a high log-likelihood) for the larger values of $k$; thus, the parametric bootstrap sheds some light on how the asymptotics of the BIC become unreliable if the number of parameters is too large. This, however, would probably not distract the researcher from declaring $k=8$ to be the optimal number of clusters here.

\subsection{Non-clustering structure}
The reasoning in the previous subsection does not take into account the way the snails data were pre-processed, though. Instead, it takes the MDS output at face value. But there is some structure in the original data, which could explain an apparent clustering in the MDS output. The presences of the species are spatially autocorrelated. If it is known that a species is present on a certain island $I$, it is more likely to find the species also on neighbouring islands than if the species is known not to be present on $I$. Furthermore, some islands host more species than others (``island attractivity'') because of factors such as their size and vegetation, and there is a certain distribution of species sizes $|{\bf x}|$. We interpret all this structure as not indicating clustering; a clustering should be interpreted as groups of species being attracted significantly to certain specific islands, and  other groups of species being attracted to other islands, which contradicts the idea that the species can be modelled as i.i.d., taking autocorrelation and island attractivity into account in the same way for all species. This involves some judgement; it is conceivable, for example, to interpret variations in island attractiveness as a consequence of real clustering rather than as a non-clustering feature of the data, which, however, does not agree with the ``biotic element''-concept in \cite{HaHe03}.

\subsection{Null model for spatial autocorrelation} \label{snulls}
The null model used here has already been used in \cite{HeHa04}, although in a different way, not connected with the number of clusters. According to ${\cal P}_0$, the species ${\bf x}_i,\ i=1,\ldots,n$ are modelled as i.i.d., having arisen from the following process. The parameters are an autocorrelation (``disjunction'') parameter $p_d$, the distribution of species sizes $P_S$ and the island attractivity distribution $P_I$. Assume that there is a neighbourhood list indicating for each pair of island whether they are neighbours or not.
\begin{enumerate}
\item Draw $n_i=|{\bf x}_i|$ from $P_S$.
\item Draw an initial island $I_1\in {\bf x}_i$ from $P_I$.
\item If $n_i>1$, for $j\in 2,\ldots,n_i$:
  \begin{enumerate}
  \item Let $N_0$ be the set of non-neighbours $N_0$ of all of $I_1,\ldots,I_{j-1}$ (not including $I_1,\ldots,I_{j-1}$). Let $N_1$ be the set of neighbours $N_1$ of any of $I_1,\ldots,I_{j-1}$ (not including $I_1,\ldots,I_{j-1}$). If neither $N_0$ nor $N_1$ is empty: 
  \item With probability $p_d$, draw an island $I_j$ from the set of non-neighbours $N_0$ of all of $I_1,\ldots,I_{j-1}$ (not including $I_1,\ldots,I_{j-1}$) according to $P_I$ conditionally on $N_0$. 
  \item Otherwise, draw $I_j$ from the set of neighbours $N_1$ of any of $I_1,\ldots,I_{j-1}$ (not including $I_1,\ldots,I_{j-1}$) according to $P_I$ conditionally on $N_1$. 
  \item If either $N_0=\emptyset$ or $N_1=\emptyset$, draw $I_j$ from the remaining non-empty set $N^*$ according to $P_I$ conditionally on $N^*$. 
  \item Put $I_j\in {\bf x}_i$.
  \end{enumerate}
\end{enumerate}

\subsection{Null model parameter estimation} \label{sests}
$P_S$ can be estimated by the empirical distribution of species sizes. $P_I$ has a straightforward empirical counterpart as well, namely choosing the probability for each island proportional to the number of species on that island. The estimation of $p_d$ is a bit more subtle. A naive estimator for $p_d$ would be
$q_d=\frac{\sum_{i=1}^n (a_i-1)}{\sum_{i=1}^n (n_i-1)}$, where $a_i$ is the number of connectivity components of the species distribution range ${\bf x}_i$. However, $q_d$ may not work very well because of situations with $N_0=\emptyset$ or $N_1=\emptyset$ in Section \ref{snulls}, and because initially separated connectivity components may grow together in the process of generating a species. We use the recommendation of \cite{HeHa04} to simulate the observable $q_d$ as a function of $p_d$ by sampling from the null model with various values of $p_d$, to fit a linear regression explaining $q_d$ from $p_d$, and then by estimating the real $p_d$ by plugging the real observed $q_d$ into the regression equation, as implemented in the ``prabclus''-package of R.  

\subsection{Parametric bootstrap}\label{spars}
Repeat $m$ times:
\begin{enumerate}
\item Estimate the null model parameters according to Section \ref{sests}.
\item Generate $n$ i.i.d. species in the bootstrap sample ${\bf X}^*$
according to the algorithmic null model in Section \ref{snulls}.
\item Compute the Kulczynski dissimilarity matrix between the species.
\item Map them onto $\mathbb{R}^4$ by classical MDS.
\item Cluster them fitting a Gaussian mixture model with a uniform noise component for $k\in K$.
\item Let $V_{qk}=V({\bf X}^*, C({\bf X}^*,k))$, $q=1,\ldots,m$, be the resulting value of $\frac{{\rm BIC}_{q}(k)-{\rm BIC}_{q}(1)}{{\rm BIC}_{q}(1)}$, where ${\rm BIC}_q(k)$ is the BIC-value for $k$ mixture components.
\end{enumerate}
The last step contains a subtle adjustment. Instead of the plain BIC, $V_{qk}$ adjusts the BIC by the BIC-value for $k=1$. The reason for this is that the parametric bootstrap test compares a clustering alternative, i.e., $k>1$, with a null model for a homogeneous population, i.e., $k=1$. A dataset with large BIC(1) can be expected to have smaller variation (as expressed by covariance matrix eigenvalues) than a dataset with smaller BIC(1), and therefore also the former dataset's BIC-values for larger $k$ will likely be larger. The adjustment corrects for this. 

\subsection{Results}
\begin{figure}
\begin{center}
  \includegraphics[width=0.5\textwidth]{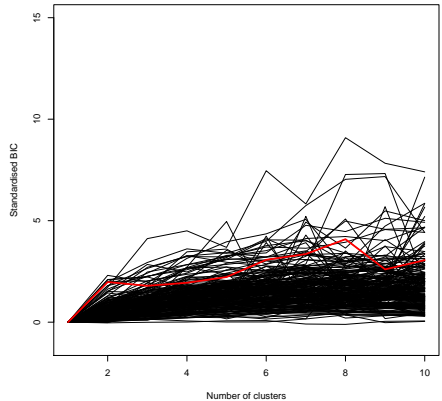}
\end{center}
\caption{($k=1$)-adjusted BIC values for Aegean islands snails data (red) and
500 bootstrap samples (black) drawn from null model with spatial autocorrelation for $k=1,\ldots,10$.}
\label{fbicnull}       
\end{figure}
The results are shown in Figure \ref{fbicnull}. The real dataset does not really stick out, but its $(k=1)$-adjusted BIC-values ($V$) are certainly among the higher ones generated by the null model. (\ref{ep}) yields $\hat p=0.0697$, so that the BIC-values of the snails data fail rather tightly to be significant evidence for real clustering at the 5\%-level. Taking into account the ad hoc nature of the model and the estimation, and the resulting anti-conservativity of the p-values (see Section \ref{sbasic}), the stronger statement that there is no evidence for clustering seems justified (by the way, using the raw BIC as $V$ produces an even higher $\hat p$). Furthermore, different from what the raw BIC-values indicated, (\ref{ek}) indicates $k=2$ as clearly better, with calibrated $V$ of 3.149, than $k=8$ (second best with calibrated $V$ of 1.971). Overall, taking into account the spatial autocorrelation of the unprocessed presence-absence data in this way changes the results quite a bit, compared to Section \ref{splains}, and can particularly explain the extent to which clustering is observed through the BIC.

\section{Concluding discussion} \label{sconc}
The present paper provides a general scheme by which cluster validation indexes can be used for testing homogeneity against a clustering alternative, and for calibrating the indexes so that their expected distribution can be taken into account for estimating the number of clusters. In the examples, it was shown that this approach can expose weaknesses of standard recommendations for how to use the validation indexes to estimate the number of clusters, and that it can detect that in some situations, in which researchers obtain a clustering, there is no evidence for any real clustering at all.

The scheme cannot be applied ``straight out of the box'', but requires the researcher to make judgements on which structural features of the data cannot be taken as indicating a ``real clustering'', and to use these for defining an appropriate null model. The examples shown here should illustrate how to go on about this task in a real situation.

The null models and parameter estimators used here have been set up in some kind of ad hoc-fashion, partly taking into account information from looking at the data. Also, parametric bootstrap will generally result in anti-conservative p-values, because the p-values are simulated from specific parameter values, and it cannot be ruled out that other parameter values can be found that are compatible with the data and yield still better values of the cluster validation index. 

Both of these issues do not affect the interpretation of non-significant outcomes, because in any case a model with certain parameter values has been found that can explain the observed clustering. This means that despite the shortcomings the interpretation is valid that there is no evidence for real clustering.

On the other hand, significant outcomes have to be interpreted with more care. They are certainly more convincing if the $V$-values for the real dataset look clearly different from those from all datasets generated by the null model in the bootstrap validity plot, rather than only just achieving $\hat p<0.05$ or 0.01. Some sensitivity analysis, i.e., running the parametric bootstrap with slightly different parameters for the null model, may give the researcher a clearer impression of how stable the significance is.

The ad hoc-character of the null models and estimators presented here may not satisfy the theoretically oriented statistician, but it is meant to encourage the data analyst to set up such models where they could be helpful even if no worked out theory is available. Apart from the direct benefit of having a test of homogeneity and a calibration of indexes for estimating the number of clusters, it is also potentially instructive to think about clustering and non-clustering structural features of the data having the task of setting up such a null model in view. Carrying out this task may give researchers a clearer idea of what kind of clusters they are looking for, and what it means, in their field, to distinguish ``clustered'' from ``homogeneous'' data.

Further research is required regarding comparing different schemes for computing p-values and estimating $k$ as mentioned in Section \ref{sbasic}. Another interesting issue is whether for estimating $k$ the index values for $k$ clusters should rather be compared with what is expected if there are $k-1$ true clusters than with a homogeneous model with only one cluster. This, however, would require the construction of probability models for all numbers of clusters. In principle one could think of fitting a mixture of null models with different parameters per cluster. In many cases, this will require considerable effort, and it can also not necessarily be taken for granted that a homogeneous null model for fitting the whole dataset is also suitable for fitting (and implicitly defining) a cluster. For example, one may want low within-cluster distances, but the null models used here are not constructed with having such an objective in mind.

{\bf Acknowledgement:} This work is supported by 
EPSRC grant EP/K033972/1.

\bibliographystyle{chicago}
\bibliography{paramboot}

\end{document}